%% file: skeleton.tex
\documentclass{PoS}

\usepackage{amsmath}
\usepackage{amssymb}
\usepackage{subfigure}

\DeclareMathAlphabet{\mathcal}{OMS}{cmsy}{m}{n}

\graphicspath{{./}}

\input{newcommands.tex}

\title{Pseudoscalar flavor-singlet mesons from lattice QCD}

\ShortTitle{Pseudoscalar flavor-singlet mesons from lattice QCD}

\author{\speaker{Konstantin Ottnad}\\
        PRISMA Cluster of Excellence and Institut f\"ur Kernphysik, Johann-Joachim-Becher-Weg 45, University of Mainz, 55099 Mainz, Germany \\
        E-mail: \email{kottnad@uni-mainz.de}}

\abstract{We investigate the masses and mixing of $\eta$, $\eta'$ mesons in the framework of twisted mass lattice QCD with $N_f=2+1+1$ dynamical quark flavors. For the first time we perform a controlled chiral and continuum extrapolation to obtain physical results. To this end, we have analyzed 17 gauge ensembles with pion masses ranging from $220\,\mathrm{MeV}$ to $500\,\mathrm{MeV}$ and at three values of the lattice spacing. This calculation became feasible through the application of a powerful variance reduction technique for the computation of quark-disconnected diagrams together with several improvements in the final analysis, including a correction of the relevant correlation functions for residual topological finite volume effects. We obtain physical results for the masses $M_{\eta}=557(11)_\mathrm{stat}(03)_{\chi\mathrm{PT}}\,\mathrm{MeV}$ and $M_{\eta'}=911(64)_\mathrm{stat}(03)_{\chi\mathrm{PT}}\,\mathrm{MeV}$, as well as the mixing angle in the quark flavor basis $\phi=38.8(2.2)_\mathrm{stat}(2.4)_{\chi\mathrm{PT}}^\circ$, in excellent agreement with other results from phenomenology. Similarly, for the physical values of the decay constant parameters we find $f_l=125(5)_\mathrm{stat}(6)_{\chi\mathrm{PT}}\,\mathrm{MeV}$ and $f_s=178(4)_\mathrm{stat}(1)_{\chi\mathrm{PT}}\,\mathrm{MeV}$. Finally, we present a test of the Veneziano Witten formula using our lattice data.}

\FullConference{The 9th International workshop on Chiral Dynamics\\
		            17-21 September 2018\\
                Durham, NC, USA}

\begin{document}

\section{Introduction} \label{sec:Introduction}
In comparison with the octet mesons the $\eta'$ meson features an exceptionally large mass close to $1\,\gev$. Naively, one expects the masses of all nine mesons to be of similar size apart from corrections due to different quark masses, as they are all rooted in the spontaneous breaking of chiral symmetry. It is only the axial anomaly and the nontrivial topological nature of quantum chromodynamics that facilitate an explanation of this peculiar feature. This becomes manifest in the famous Veneziano-Witten formula \cite{Witten:1979vv,Veneziano:1979ec} which for non-vanishing quark masses reads
\begin{equation}
 \Metap^2 + \Meta^2 - 2M_K^2 = \frac{4N_f}{f_0^2} \chi^\mathrm{YM}_\infty \,.
 \label{eq:VW_formula}
\end{equation}
It relates the $\eta'$, $\eta$, kaon masses and the flavor-singlet decay constant parameter $f_0$ in dynamical QCD with the topological susceptibility $\chi^\mathrm{YM}_\infty$ in pure Yang-Mills theory. From a modern point of view the formula is obtained as a leading order results in chiral perturbation theory ($\chi$PT) for $N_f=3$ quark flavors employing a combined power counting scheme in quark masses, (small) momenta and $1/N_c$, where $N_c=3$ is the number of colors. In particular, the formula implies a non-vanishing mass for the $\eta'$ meson even in the chiral limit. It is only in the t'Hooft limit ($N_c\rightarrow \infty$, while $N_f$ and $g^2N_c$ are kept fixed, with $g$ denoting the gauge coupling) that the $\eta'$ becomes massless and the $U(1)_A$ symmetry is restored. \par

While the Veneziano-Witten formula provides a qualitative explanation for the large mass of the $\eta'$, a direct, quantitative study requires non-perturbative methods. In principle, lattice QCD allows to compute meson masses including the $\eta'$ as well as decay constants and mixing parameters from first principles. Furthermore, it is also possible to directly compute $\chi^\mathrm{YM}_\infty$ using lattice methods. However, the required calculations for the $\eta$ and $\eta'$ turn out to be very demanding from a computational point of view. This is due to the essential contribution of quark-disconnected diagrams in the relevant two-point functions and the dependence of the observables on an accurate sampling of the topological properties of the theory. \par

In this study we use lattice QCD to compute masses for the $\eta$ and $\eta'$ meson as well as mixing parameters $\phi$, $f_l$ and $f_s$ in the quark flavor basis. The latter allows us to obtain also an estimate for $f_0$ which is defined in the octet-singlet basis. The proceedings contribution is organized as follows: In Sec.~\ref{sec:Lattice setup} we give an overview on the lattice action and the gauge ensembles used in this study. The construction of the required correlation functions and some essential analysis methods are described in Sec.~\ref{sec:Correlation functions and improvement for topological effects}, including a brief discussion of an improvement for topological effects in the $\eta$, $\eta'$ correlators. Results for the masses and mixing parameters from combined chiral and continuum extrapolations are presented in Sec.~\ref{sec: Masses and mixing} together with the respective. Finally, in Sec.~\ref{sec:Test of the Veneziano-Witten formula} we perform a test of the Veneziano-Witten Formula using our lattice results. \par

Most of the methods and results discussed in this proceedings contribution have previously been published in Refs.~\cite{Ottnad:2012fv,Michael:2013gka,Ottnad:2015hva,Cichy:2015jra,Ottnad:2017bjt}. The actual numerical results are taken from the improved and most comprehensive analysis in Ref.~\cite{Ottnad:2017bjt}. Since these results feature several systematic improvements compared to earlier publications we have included an update of the fermionic part of the analysis for the Veneziano-Witten formula in Ref.~\cite{Cichy:2015jra} based on this improved data. \par

\section{Lattice setup} \label{sec:Lattice setup}
The numerical calculations for the analysis of $\eta$ and $\eta'$ masses and mixing parameters have been performed on a set of 17 gauge ensembles provided by the European Twisted Mass Collaboration (ETMC) \cite{Baron:2010bv,Baron:2010th,Baron:2011sf}. These ensembles have been generated with $N_f=2+1+1$ dynamical quark flavors of twisted mass Wilson fermions at maximal twist. An overview of the ensembles used in this study is given Table~\ref{tab:ensembles}. \par

\begin{table}[t!]
 \centering
 \begin{tabular*}{\textwidth}{@{\extracolsep{\fill}}lllllllrll}
  \hline\hline
  ensemble  & $\beta$ & T/a & L/a & $a\mu_l$ & $a\mu_\sigma$ & $a\mu_\delta$ & $N_\mathrm{conf}$ & $aM_\pi$ & $aM_K$ \\
  \hline\hline
  A30.32   & 1.90 & 64 & 32 & 0.0030 & 0.150  & 0.190  & 1363 & 0.12384(53) & 0.2511(07) \\
  A40.32   & 1.90 & 64 & 32 & 0.0040 & 0.150  & 0.190  &  863 & 0.14128(26) & 0.2569(07) \\
  A40.24   & 1.90 & 48 & 24 & 0.0040 & 0.150  & 0.190  & 1877 & 0.14520(40) & 0.2590(09) \\
  A60.24   & 1.90 & 48 & 24 & 0.0060 & 0.150  & 0.190  & 1248 & 0.17316(38) & 0.2663(11) \\
  A80.24   & 1.90 & 48 & 24 & 0.0080 & 0.150  & 0.190  & 2449 & 0.19922(30) & 0.2779(08) \\
  A100.24  & 1.90 & 48 & 24 & 0.0100 & 0.150  & 0.190  & 2489 & 0.22161(35) & 0.2878(08) \\
  \hline                                                                                
  A80.24s  & 1.90 & 48 & 24 & 0.0080 & 0.150  & 0.197  & 2514 & 0.19895(42) & 0.2550(05) \\
  A100.24s & 1.90 & 48 & 24 & 0.0100 & 0.150  & 0.197  & 2312 & 0.22207(27) & 0.2655(11) \\
  \hline                                                                                
  B25.32   & 1.95 & 64 & 32 & 0.0025 & 0.135  & 0.170  & 1467 & 0.10708(32) & 0.2130(06) \\
  B35.32   & 1.95 & 64 & 32 & 0.0035 & 0.135  & 0.170  & 1251 & 0.12530(28) & 0.2181(06) \\
  B55.32   & 1.95 & 64 & 32 & 0.0055 & 0.135  & 0.170  & 4996 & 0.15567(17) & 0.2288(02) \\
  B75.32   & 1.95 & 64 & 32 & 0.0075 & 0.135  & 0.170  &  922 & 0.18082(30) & 0.2378(07) \\
  B85.24   & 1.95 & 48 & 24 & 0.0085 & 0.135  & 0.170  &  573 & 0.19299(58) & 0.2459(26) \\
  \hline                                                                                
  D15.48   & 2.10 & 96 & 48 & 0.0015 & 0.120  & 0.1385 & 1034 & 0.06912(30) & 0.1691(12) \\
  D20.48   & 2.10 & 96 & 48 & 0.0020 & 0.120  & 0.1385 &  429 & 0.07870(26) & 0.1732(03) \\
  D30.48   & 2.10 & 96 & 48 & 0.0030 & 0.120  & 0.1385 &  458 & 0.09788(29) & 0.1774(04) \\
  D45.32sc & 2.10 & 64 & 32 & 0.0045 & 0.0937 & 0.1077 & 1074 & 0.11847(54) & 0.1747(04) \\
  \hline\hline
 \end{tabular*}
 \caption{Overview of ensembles and input parameters. In addition, we give the number of gauge configurations $N_\mathrm{conf}$ used as well as pion and kaon masses in lattice units with their respective statistical errors.}
 \label{tab:ensembles}
\end{table}

In the gauge sector the Iwasaki action \cite{Iwasaki:1985we,Iwasaki:1996sn} has been used in the generation of configurations
\begin{equation}
 S_G[U]=\frac{\beta}{3}\sum_x\left(b_0\sum_{\genfrac{}{}{0pt}{}{\mu,\nu=1}{1\leq\mu<\nu}}^4\mathrm{Re}\,\tr \left(1-P^{1\times 1}_{x;\mu\nu}\right)+b_1\sum_{\genfrac{}{}{0pt}{}{\mu,\nu=1}{\mu \neq \nu}}^4\mathrm{Re}\, \tr \left(1-P^{1\times 2}_{x;\mu\nu}\right)\right)\,,
 \label{eq:iwasaki_action}
\end{equation}
where $b_1=-0.331$ and $b_0=1-8b_1$. $P^{1\times 1}_{x;\mu\nu}$ and $P^{1\times 2}_{x;\mu\nu}$ denote quadratic and rectangular Wilson loops consisting of gauge links. The fermionic part of the action with a mass-degenerate, light quark doublet $\chi_l=(\chi_u,\chi_d)^T$ reads \cite{Frezzotti:2000nk,Frezzotti:2003ni,Frezzotti:2004wz}
\begin{equation}
  S_l[\chi_l,\chibar_l,U]=a^4 \sum_x \chibar_l(x)(D_W[U]+m_0+i\mu_l\gamma_5\tau_3)\chi_l(x)\,,
 \label{eq:light_action}
\end{equation}
while for a non-degenerate, heavy quark doublet $\chi_h=(\chi_c,\chi_s)^T$ it takes the form \cite{Frezzotti:2004wz,Frezzotti:2003xj}
\begin{equation}
S_h[\chi_h,\chibar_h,U]=a^4 \sum_x\chibar_h(x)(D_W[U]+m_0+i\mu_\sigma\gamma_5\tau_1+\mu_\delta\tau_3)\chi_h(x)\,,
 \label{eq:heavy_action}
\end{equation}
where $\tau^i$, $i=1,2,3$ denote Pauli matrices acting in flavor space. The massless Wilson Dirac operator $D_W=\frac{1}{2}(\gamma_\mu(\nabla_\mu+\nabla^\star)-a\nabla^\star_\mu\nabla_\mu)$ depends implicitly on the gauge links $U$. Both of doublet fields in the twisted mass basis $\chi_l$ and $\chi_h$ are related to doublet fields in the physical basis $\psi_l$, $\psi_h$ via chiral rotations. The bare strange and charm quark masses are related to the bare input parameters $\mu_\delta$, $\mu_\sigma$ by
\begin{equation}
 \mu_{c,s}=\mu_\sigma\pm Z\mu_\delta \,,
 \label{eq:heavy_quark_masses}
\end{equation}
while the renormalized quark masses picks up an additional factor of non-singlet $1/Z_P$
\begin{equation}
 \mu_{c,s}^r = Z_P^{-1} \mu_\sigma\pm Z_S^{-1} \mu_\delta \,,
 \label{eq:heavy_quark_masses_renormalized}
\end{equation}
similar to the light bare quark mass, i.e. $\mu_l^r= \mu_l / Z_P$. \par

\begin{table}[t!]
 \centering
 \begin{tabular*}{.5\textwidth}{@{\extracolsep{\fill}}lllll}
 \hline\hline
 $\beta$  & $r_0/a$ & $a\,[\mathrm{fm}]$ & Z \\
 \hline\hline
 1.90 & 5.31(8) & 0.0885(36) & 0.651(06) \\
 1.95 & 5.77(6) & 0.0815(30) & 0.666(04) \\
 2.10 & 7.60(8) & 0.0619(18) & 0.727(03) \\
 \hline\hline
 \end{tabular*}
 \caption{Values of $r_0/a$, $a$ and $Z$ corresponding to the three values of $\beta$ as given in Ref.~\cite{Carrasco:2014cwa}. The values for $Z$ are taken from method M2 in this publication.}
 \label{tab:beta_r0_a_Z}
\end{table}

The ensembles used in this work cover light quark masses corresponding to a pion mass range of $220\mev$ to $500\mev$ and three values of the lattice spacing $a$ as listed in Table~\ref{tab:beta_r0_a_Z}. In the same table we have included the ratio of pseudoscalar and scalar flavor non-singlet renormalization factors $Z=Z_P/Z_S$. The ratio $Z$ enters the kaon decay constant $f_K$ and the calculation of decay constant parameters related to the mixing of the $\eta$ and $\eta'$ mesons. Besides, Table~\ref{tab:beta_r0_a_Z} contains chirally extrapolated values of the Sommer scale $r_0/a$ at each lattice spacing as obtained in Ref.~\cite{Carrasco:2014cwa}. These values are used to form dimensionless quantities in the chiral and continuum extrapolations. The physical value $r_0=0.474(14)_\stat$ determined in the same study is used to set the scale in our simulations. For more detailed numerical results for meson masses (including  $\Meta$ and $\Metap$) and further observables such as decay constants and mixing parameters we refer to the tables in Ref.~\cite{Ottnad:2017bjt}. \par

While the bare strange and charm quark mass parameters have generally been chosen constant for each value of the lattice spacing, they have been varied on a few ensembles (A80.24s, A100.24s and D45.32sc), which can be inferred from the values for $\mu_\sigma$, $\mu_\delta$ and the resulting kaon masses on these ensembles in Tab.~\ref{tab:ensembles}. This allows us to explicitly disentangle effects related to changes in the strange quark mass and gives a better handle on the corrsponding chiral extrapolations. \par

\section{Correlation functions and improvement for topological effects} \label{sec:Correlation functions and improvement for topological effects}
In the physical basis we consider three local pseudoscalar operators build from the doublet fields $\psi_{l}$ and $\psi_{h}$
\begin{equation}
 \mathcal{P}_l^{0,\phys}(x) = \frac{1}{\sqrt{2}} \bar{\psi}_l(x) i \g{5} \psi_l(x) \quad \text{and} \quad \mathcal{P}_h^{\pm,\phys}(x) = \bar{\psi}_h(x) i \g{5} \frac{1\pm\tau^3}{2}\psi_h(x) \,. \label{eq:ops_phys}
\end{equation}
Since the charm quark has essentially no overlap with the $\eta$ and $\eta'$ states \cite{Michael:2013gka} we drop the operator related to the charm quark from the subsequent analysis. In the twisted mass basis at maximal twist and including renormalization up to a factor of $Z_P$ the resulting two operators for light and strange quarks read
\begin{equation}
 \tilde{\mathcal{S}}_l^{\tm}(x) = \frac{1}{\sqrt{2}} Z^{-1} \bar{\chi}_l(x) \tau^3 \chi_l(x) \quad \text{and} \quad \tilde{\mathcal{P}}_s^{\tm}(x) = \frac{1}{2} \bar{\chi}_h(x) \l(-Z^{-1}\tau^1 - i \g{5} \tau^3\r) \chi_h(x) \,. \label{eq:opd_Z_only}
\end{equation}
These operators give rise to a $2\times2$ correlation function matrix
\begin{equation}
 \tilde{\mathcal{C}}(t) = \l(\begin{array}{cc}
  \bigl<\tilde{\mathcal{S}}_l^{\tm}(t)\tilde{\mathcal{S}}_l^{\tm}(0)\bigr> &
  \bigl<\tilde{\mathcal{S}}_l^{\tm}(t)\tilde{\mathcal{P}}_s^{\tm}(0)\bigr> \\
  \bigl<\tilde{\mathcal{P}}_s^{\tm}(t)\tilde{\mathcal{S}}_l^{\tm}(0)\bigr> &
  \bigl<\tilde{\mathcal{P}}_s^{\tm}(t)\tilde{\mathcal{P}}_s^{\tm}(0)\bigr> \end{array}\r)
 \label{eq:corr_matrix}
\end{equation}
that is renormalized up to a factor of $Z_P^2$, i.e. $\mathcal{C}^r(t)=Z_P^2\tilde{\mathcal{C}}(t)$. This is sufficient for the calculation of mixing parameters as factors of $Z_P$ are always canceled by matching quark mass factors. \par

All the two-point functions in Eq.~\ref{eq:corr_matrix} contain quark-disconnected pieces, which are computationally very demanding as they require all-to-all propagators. In fact, the off-diagonal matrix elements receive no quark-connected contribution at all. Therefore, it is crucial to achieve a reasonable signal-to-noise ratio for the quark-disconnected contributions. To this end, we employ stochastic volume sources with typically 24 or 32 noise vector per configuration, depending on the lattice volume. For the light quark sector a particularly powerful variance reduction method \cite{Jansen:2008wv} is available due to the use of the twisted mass formulation of lattice QCD, while for the strange and charm sector we apply the hopping parameter expansion noise reduction technique. \par

In order to further reduce statistical errors we replace the quark-connected pieces by their respective (fitted) ground state contributions, which has first been proposed in Ref.~\cite{Neff:2001zr}. Applying this modification before solving the generalized eigenvalue problem (GEVP) 
\begin{equation}
 \tilde{\mathcal{C}}(t) v^{n}(t,t_0) = \lambda^{n}(t,t_0) \tilde{\mathcal{C}}(t_0) v^{n}(t,t_0)
 \label{eq:GEVP}
\end{equation}
allows to extract masses and amplitudes for the $\eta$ and $\eta'$ states from the resulting principal correlators $\lambda^{\eta,\eta'}(t,t_0)$ and eigenvectors $v^{\eta,\eta'}(t,t_0)$ starting from timeslices as early as $t=2a$, leading to a significant improvement in the respective signal-to-noise ratios. \par

\begin{figure}[t]
 \centering
 \includegraphics[totalheight=0.222\textheight]{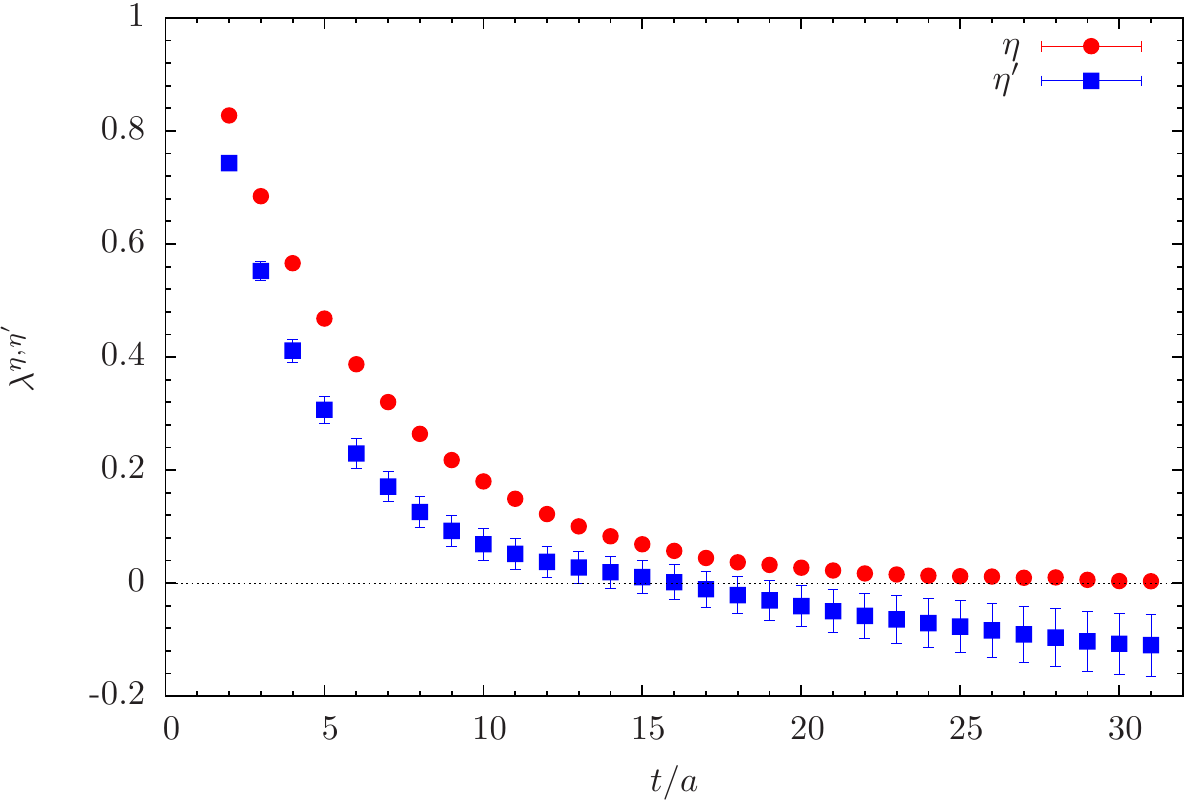} \hspace{0.5em}
 \includegraphics[totalheight=0.222\textheight]{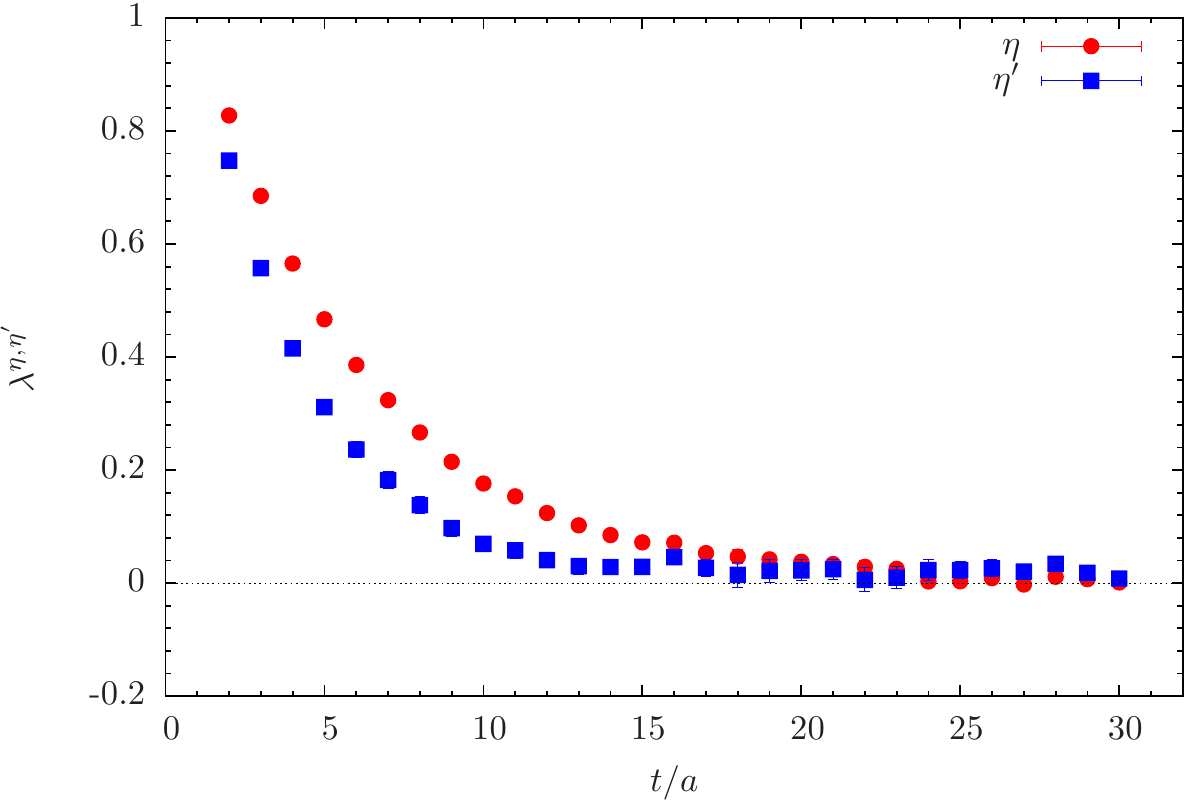}
 \caption{Principal correlators for the $\eta$ and $\eta'$ states obtained from solving the GEVP in Eq.~(\protect\ref{eq:GEVP}) for the correlation function matrix in Eq.~(\protect\ref{eq:corr_matrix}) (left panel) and the discrete time derivative correlation function matrix in Eq.~(\protect\ref{eq:corr_matrix_timeshift}) (right panel).}
\label{fig:D45_eigenvalues}
\end{figure}

A final improvement concerns residual effects of imperfectly sampling topology in finite volume, which affects the large-$t$ behavior of the quark-disconnected contribution $C_\mathrm{2pt}^\mathrm{disc}(t)$ to the pseudoscalar flavor-singlet correlation functions \cite{Aoki:2007ka,Bali:2014pva}. At fixed topological charge $Q$ the leading correction in the $1/V$ expansion at large Euclidean time separations reads
\begin{equation}
 C_\mathrm{2pt}^\mathrm{disc}(t) \sim \frac{a^5}{T}\l(\chi_\infty - \frac{Q^2}{V} + \frac{c_4}{2V\chi_\infty}\r) \,,
 \label{eq:topFV}
\end{equation}
where $c_4$ denotes the kurtosis of the topological charge distribution. Although all our ensembles exhibit zero topological charge within errors in the gauge average, one still expects that the imperfectly sampled topological charge distribution at finite statistics leads to deviations from the infinite volume result. While we observe only on a few ensembles a clearly non-zero shift of the correlation functions / principal correlators, these deviations can still introduce systematic effects for e.g. the $\eta'$ mass.  An example for a non-zero shift is shown in the left panel of Fig.~\ref{fig:D45_eigenvalues} for the ensemble D45.32sc which exhibits the smallest physical volume of all our ensembles. In order to remove this constant term, we replace $\tilde{\mathcal{C}}(t)$ by the discrete time derivative
\begin{equation}
 \tilde{\mathcal{C}}(t) \rightarrow \tilde{\mathcal{C}}(t) - \tilde{\mathcal{C}}(t+\Delta t) \,,
 \label{eq:corr_matrix_timeshift}
\end{equation}
for which the GEVP analysis can be performed in the usual way. Since our simulations are performed with periodic bounday conditions in time, the only difference is a change from a $cosh$-like to a $sinh$-like behavior in $t$ for the resulting principal correlators when replacing the original correlation functions by their discrete time derivatives. The result of this procedure is shown in the right panel of Fig.~\ref{fig:D45_eigenvalues}, where the large-$t$ behavior of the $\eta'$ principal correlator is clearly improved compared to the original one in the left panel. \par

\section{Masses and mixing} \label{sec: Masses and mixing}
\begin{figure}[t]
 \centering
 \includegraphics[totalheight=0.215\textheight]{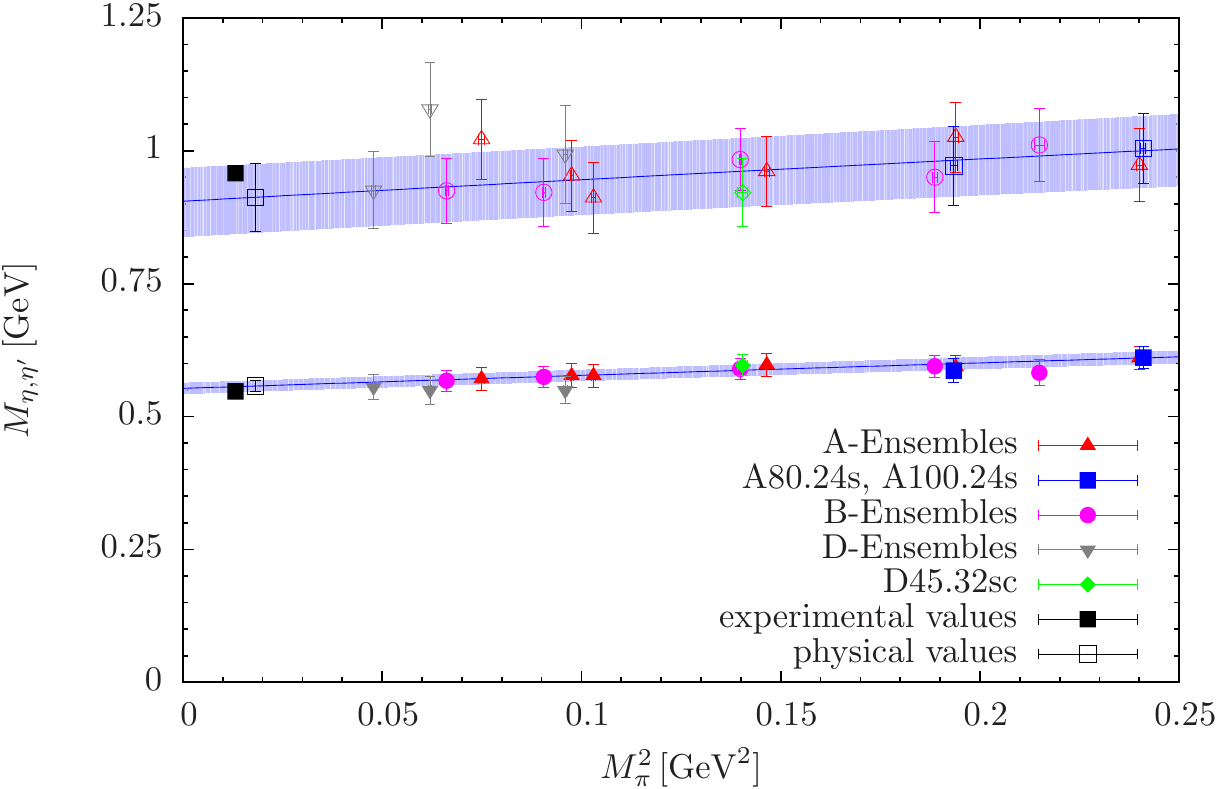} \hspace{0.5em}
 \includegraphics[totalheight=0.215\textheight]{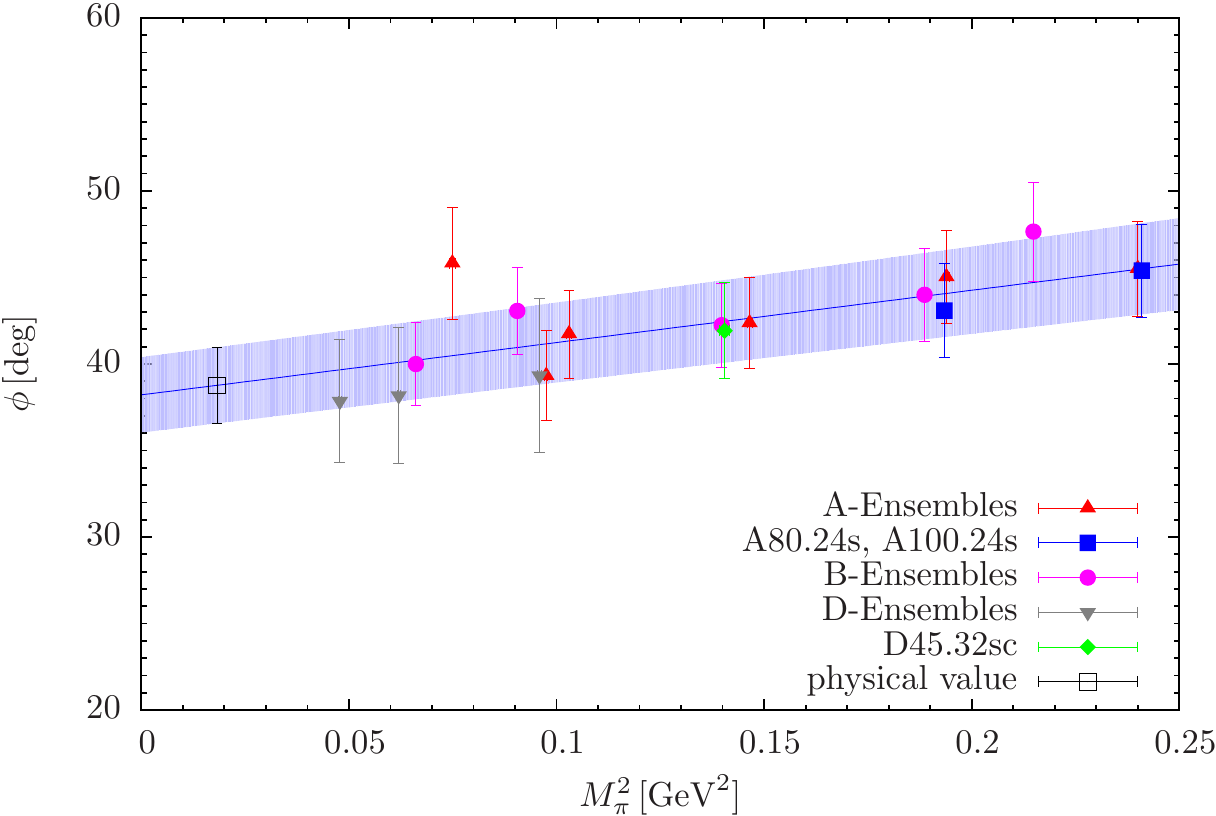}
 \caption{Combined chiral and continuum extrapolation for $\Meta$, $\Metap$ (left panel) and the mixing angle $\phi$ (right panel). Data are corrected for $\mathcal{O}(a^2)$ lattice artifacts and to physical $m_s$ using the parameters obtained from the leading order chiral fits. Point errors are highly correlated due to this correction.}
 \label{fig:masses_and_mixing}
\end{figure}

In the left panel of Fig.~\ref{fig:masses_and_mixing} we show the combined chiral and continuum extrapolation for $\Meta$ and $\Metap$ together with the lattice data that have been corrected to physical $m_s$ and for $\mathcal{O}(a^2)$ lattice artifacts. The fit model used for all our observables $O$ is inspired by leading order $\chi$PT including the leading lattice artifact of $\mathcal{O}(a^2)$
\begin{equation}
  \l(r_0^n O[r_0^2\Delta_l, r_0^2\Delta_s, (a/r_0)^2]\r)^m = (r_0^n \chiral{O})^m + \sum_{i=l,s} L_i \cdot r_0^2 \Delta_i + L_\beta \cdot \l(\frac{a}{r_0}\r)^2
  \label{eq:fit_ansatz}
\end{equation}
where $n$ is an integer chosen s.t. $r_0^n O$ is dimensionless and $m$ denotes the power of $O$ in the chiral expansion. The quantities $\Delta_l = M_\pi^2 = 2 B_0 m_l + \mathcal{O}(m^2)$, $\Delta_s = 2 M_K^2 - M_\pi^2 = 2 B_0 m_s + \mathcal{O}(m^2)$ are used as leading order proxies for the light and strange quark mass, respectively. The terms proportional to the free fit parameters $L_{l,s,\beta}$ are always present, while a constant $\chiral{O}$ appears as a free fit parameter only if $O$ has a non-trivial value in the chiral limit (e.g. $M_\eta'$). Otherwise it is set to its analytically known value (e.g. zero for $M_\eta$). Statistical errors are determined using the blocked bootstrap method with 10000 samples. Block lengths are chosen in such a way that the effective length of each block corresponds to at least 20 HMC trajectories, which has been found to be sufficient to account for autocorrelations. We assign a systematic error (labeled ``$\chi$PT'') to our results from the difference of the results fitting the full set of data and excluding data with $M_\pi>390\mev$, which should reflect the uncertainty related to the chiral extrapolation. The final results for the masses at the physical point are found to be in good agreement with experiment
\begin{equation}
 M_{\eta,\mathrm{phys}}  = 557(11)_\mathrm{stat}(03)_{\chi PT}\,\mathrm{MeV} \quad \text{and} \quad M_{\eta',\mathrm{phys}} = 911(64)_\mathrm{stat}(03)_{\chi PT}\,\mathrm{MeV} \,. \label{eq:M}
\end{equation}
Alternatively, for the $\eta$ meson it is possible to consider the ratio $M_\eta/M_K$ which exhibits a much milder $m_s$-dependence. At the physical point we find $(M_\eta/M_K)_\phys = 1.114(31)_\stat$, corresponding to $M_{\eta,\phys}=554(15)_\stat\mev$ which agrees well with the direct extrapolation. \par

In general, decay constants and mixing parameters are defined from axial vector matrix elements. However, due to an unfavorable signal-to-noise ratio for axial vector interpolating operators we restrict ourselves to the computation of (renormalized) pseudoscalar matrix elements, which are given by
\begin{equation}
 h^{P,\tm,r}_l = \mu_l \l<0\r| \tilde{\mathcal{S}}_l^{\tm} \l|P\r> \quad \text{and} \quad h^{P,\tm,r}_s = \mu_s \l<0\r| \tilde{\mathcal{P}}_s^{\tm} \l|P\r> \,,
 \label{eq:PS_matrix_elements}
\end{equation}
for $P=\eta,\eta'$. Note that this is where the previously mentioned cancellation of non-singlet $Z_P$-factors enters. For the determination of mixing parameters we consider the so-called FKS scheme in the quark flavor basis with a single mixing angle $\phi$, neglecting OZI-suppressed terms. In this scheme the following leading order $\chi$PT relation between pseudoscalar matrix elements and the usual mixing parameters holds \cite{Feldmann:1999uf}
\begin{equation}
 \l(\begin{array}{cc}
  h^\eta_l    & h^\eta_s \\
  h^{\eta'}_l & h^{\eta'}_s
 \end{array}\r) = \l(\begin{array}{cc}
  \cos\phi  & -\sin\phi \\
  \sin\phi  & \cos\phi
 \end{array}\r) \diag\l(M_\pi^2 f_l, \l(2M_K^2-M_\pi^2\r) f_s\r) \,.
\label{eq:pseudoscalar_mixing}
\end{equation}
which allows to extract the mixing angle $\phi$ and the decay constant parameters $f_l$ and $f_s$ from our lattice data. For further details we refer to Ref.~\cite{Ottnad:2017bjt}.

\begin{figure}[t]
 \centering
 \includegraphics[totalheight=0.212\textheight]{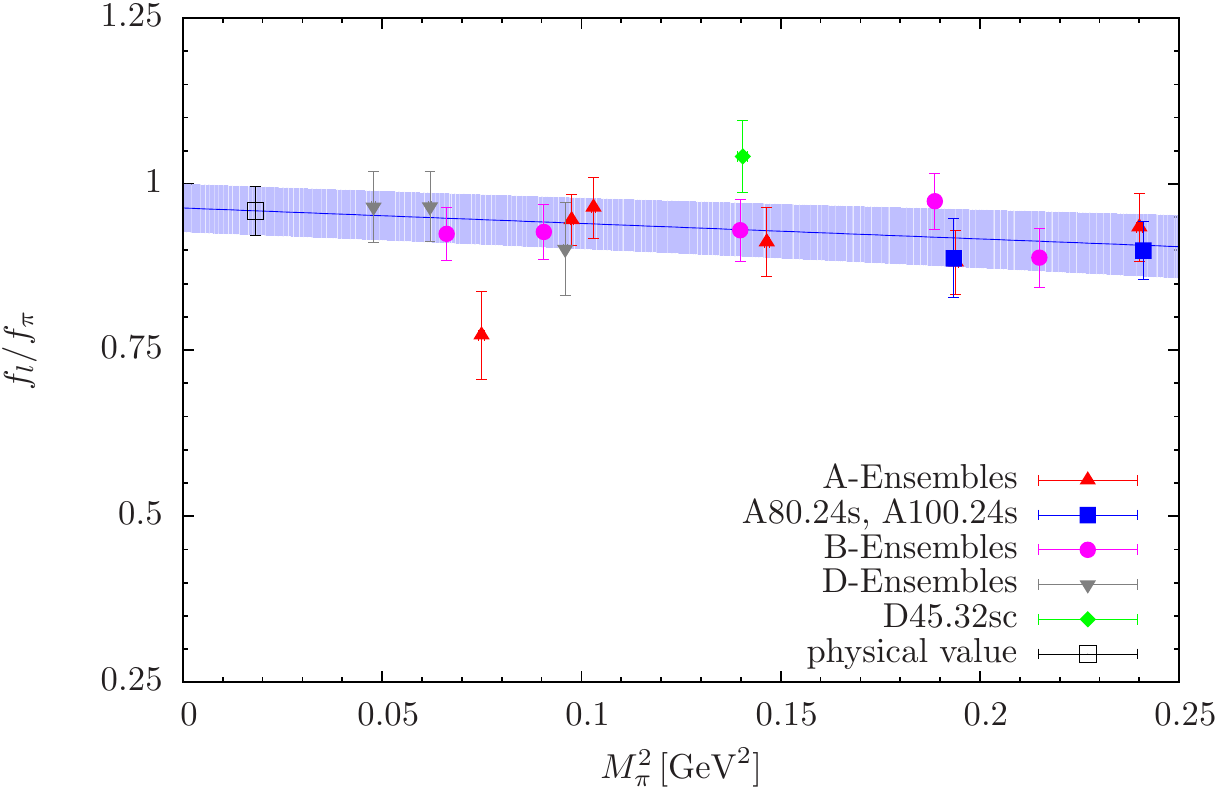} \hspace{0.5em}
 \includegraphics[totalheight=0.212\textheight]{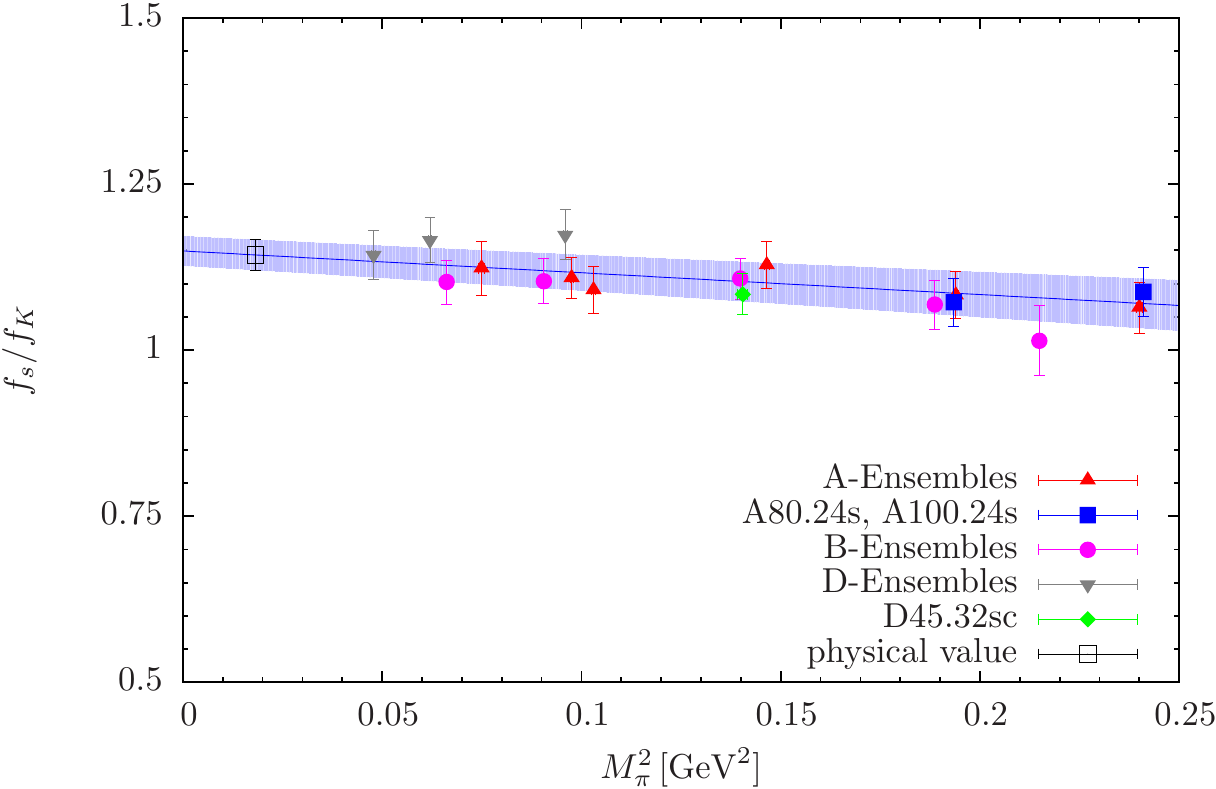}
 \caption{Combined chiral and continuum extrapolation for $f_l/f_\pi$ (left panel) and $f_s/f_K$ (right panel). Data are corrected for $\mathcal{O}(a^2)$ lattice artifacts and to physical $m_s$ using the parameters obtained from the leading order chiral fits. Point errors are highly correlated due to this correction.}
 \label{fig:decay_constants}
\end{figure}

The right panel of Fig.~\ref{fig:masses_and_mixing} shows the combined chiral and continuum extrapolation for $\phi$. In the chiral limit one has ideal mixing, i.e. $\chiral{\phi}=\arctan\sqrt{2}$,  and at the physical point we find
\begin{equation}
 \phi_\phys = 38.8^\circ(2.2)_\stat (2.4)_{\chi PT} \,,
 \label{eq:phi}
\end{equation}
which does not depend on renormalization and is in excellent agreement with recent results from phenomenology \cite{Escribano:2013kba,Escribano:2015nra,Escribano:2015yup}. For the decay constant parameters a direct extrapolation turns out difficult due to large quark mass corrections and lattice artifacts. Therefore, we consider the ratios $f_l/f_\pi$ and $f_s/f_K$ which receive milder corrections. The results from fitting the model in Eq.~(\ref{eq:fit_ansatz}) to our lattice data for these ratios is shown in Fig.~\ref{fig:decay_constants}. At the physical point we obtain
\begin{equation}
  (f_l/f_\pi)_\phys = 0.960(37)_\stat(46)_{\chi PT} \quad \text{and} \quad (f_s/f_\mathrm{K})_\phys = 1.143(23)_\stat(04)_{\chi PT} \,.  \label{eq:f_ratios}
\end{equation}
Using the experimental values $f_{\pi}^{\experiment}=130.50\mev$ and $f_{\mathrm{K}}^{\experiment}=155.72\mev$ \cite{Patrignani:2016xqp} this leads to
\begin{equation}
 f_{l,\phys} = 125(5)_\stat(6)_{\chi PT} \mev \quad \text{and} \quad f_{s,\phys} = 178(4)_\stat(1)_{\chi PT} \mev \,. \label{eq:fs}
\end{equation}

\section{Test of the Veneziano-Witten formula} \label{sec:Test of the Veneziano-Witten formula}
As mentioned before, all the quantities entering the Veneziano-Witten formula in Eq.~\ref{eq:VW_formula} can be computed in lattice QCD. For the fermionic observables the only remaining piece is the singlet decay constant parameter $f_0$ which is defined in the octet-singlet basis instead of the quark flavor basis. Following the strategy in Ref.~\cite{Cichy:2015jra}, one can relate $f_0$ to $f_l$, $f_s$, $f_\pi$ and $f_K$ to leading order in $\chi$PT. We consider three possible relations that differ by higher order terms in $\chi$PT and residual lattice artifacts
\begin{align}
 f_0^2 &= -7/6 f_\pi^2 + 2/3 f_K^2 + 3/2 f_l^2 \,,     \label{eq:f_0_D1} \\
 f_0^2 &= +1/3 f_\pi^2 - 4/3 f_K^2 + f_l^2 + f_s^2 \,, \label{eq:f_0_D2} \\
 f_0^2 &= +10/3 f_\pi^2 - 16/3 f_K^2 + 3 f_s^2 \,.     \label{eq:f_0_D3}
\end{align}
Plugging the physical values for $f_\pi$, $f_K$, $f_l$ and $f_s$ in these three definitions for $f_0$, we arrive at physical values for $f_0$, i.e. $f_{0,\phys} = 141(06)_\stat \mev$, $f_{0,\phys} = 144(07)_\stat \mev$ and $f_{0,\phys} = 149(13)_\stat \mev$, respectively. The spread in the values is of similar size as the statistical errors, which gives a hint that at the current statistical precision higher order corrections are not yet the dominating source of error. We remark that the rather different statistical errors are caused by (anti-) correlations between the data for the individual quantities that enter in different linear combinations in the three definitions of $f_0$, leading to cancellations (or enhancements) of fluctuations. \par

Combining all our results for the fermionic quantities contributing in Eq.~(\ref{eq:VW_formula}) and taking the weighted average over the three different values for $f_0$ we obtain the following results for our dynamical simulations
\begin{equation}
 r_0^4 \chi^\mathrm{dyn}_\infty = 0.037(7)_{\stat + \sys} \,,
\end{equation}
where the error reflects both, statistical and systematic effects due to higher order corrections. This new result is somewhat smaller than the one obtained previously in Ref.~\cite{Cichy:2015jra}, i.e. $r_0^4\chi_\infty^\mathrm{dyn}=0.047(03)_\stat(11)_\sys$. However, the present analysis supersedes the one from 2015 due to several systematic improvements. In particular, we have now been able to perform combined chiral and continuum extrapolations for all the individual observables. This has become feasible due to an increase of statistics on some of the ensembles and the inclusion of an additional ensemble at the finest lattice spacing (D20.48). Besides, the old analysis did not include the aforementioned correction for residual topological effects in the correlations functions. \par

\begin{figure}[t]
 \centering
 \includegraphics[totalheight=0.156\textheight]{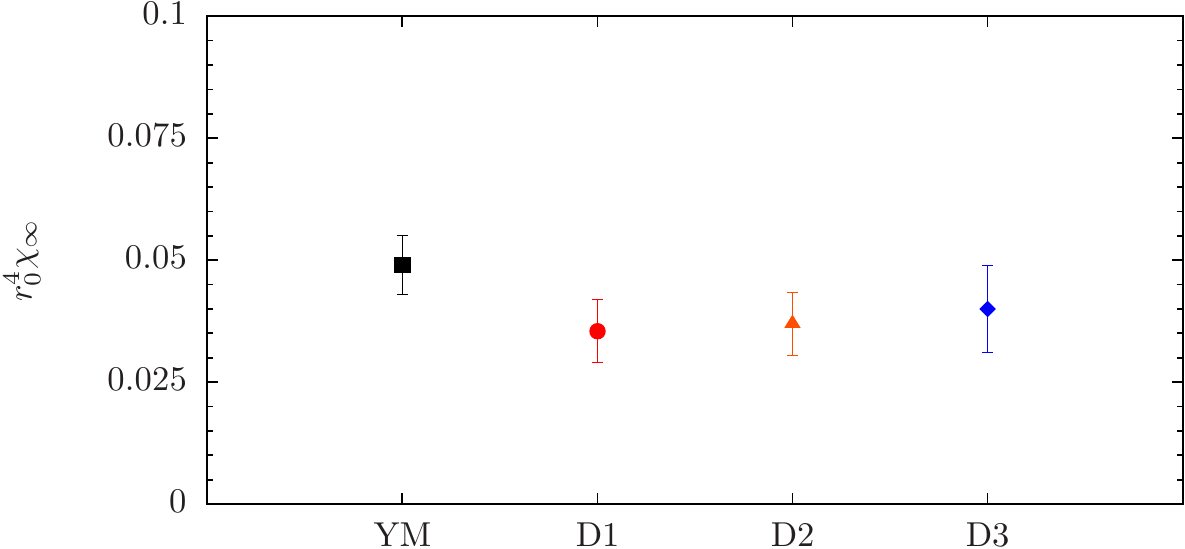}\hspace{0.5em}
 \includegraphics[totalheight=0.156\textheight]{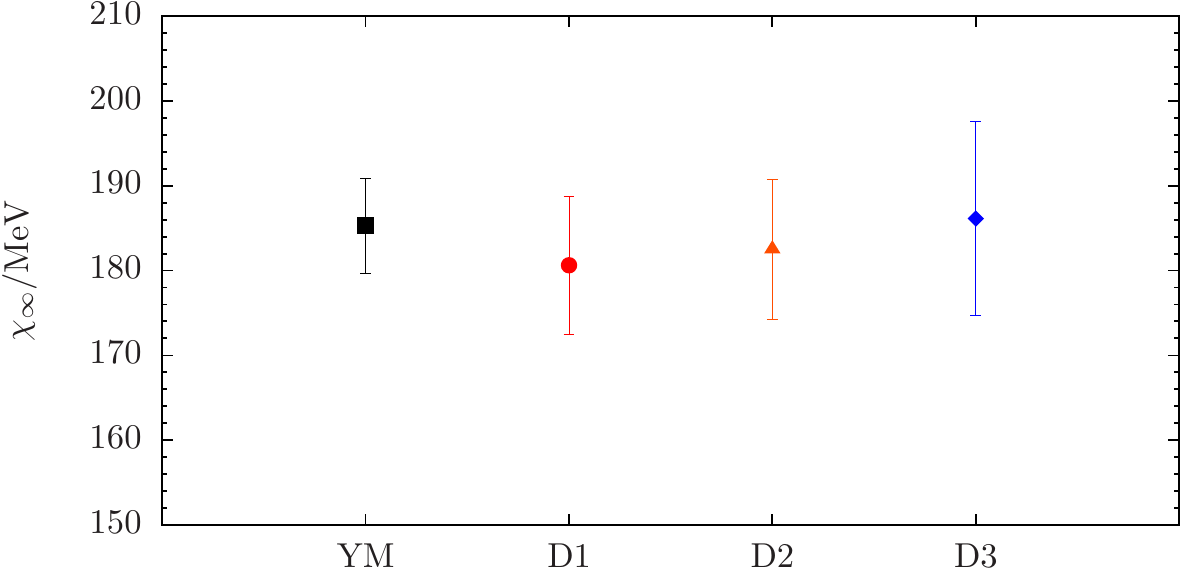}
 \caption{Comparison of results for the topological susceptibility in Yang-Mills theory (YM) and from our dynamical simulations using the three different definitions (D1, D2 and D3) in Eqs.~(\protect\ref{eq:f_0_D1}--\protect\ref{eq:f_0_D3}) for computing $f_0$. Left panel: Results in units of $r_0$. Right panel: Results in physical units; see text.}
 \label{fig:chi0}
\end{figure}

For the quenched topological susceptibility $\chi^\mathrm{YM}_\infty$ we quote the result from Ref.~\cite{Cichy:2015jra} which has been obtained from the spectral projector method and taking the continuum limit over four values of the lattice spacing. A comparison of this result with our individual results for the three definitions of $f_0$ is shown in the left panel of Fig.~\ref{fig:chi0}. However, in general there is no reason to assume that the value of $r_0$ is the same in the quenched and the dynamical theory. Assuming the standard value of $r_0^\mathrm{YM}=0.5 \,\mathrm{fm}$ for the quenched calculation we can perform a comparison in physical units, which is shown in the right panel of Fig.~\ref{fig:chi0}. We find surprisingly good agreement between the value computed in the quenched theory in Ref.~\cite{Cichy:2015jra} and the ones obtained in the dynamical theory, i.e.
\begin{equation}
 \chi^\mathrm{YM}_\infty = (185.3(5.6)_{\stat + \sys} \mev)^4 \quad \text{and} \quad \chi^\mathrm{dyn}_\infty = (182.6(8.3)_{\stat + \sys} \mev)^4 \,.
\end{equation}
This confirms once more the validity of the formula and that the $\eta'$ receives a large contribution to its mass due to the axial anomaly, which does not vanish in the chiral limit. For the two-flavor theory this result has recently been further corroborated by the study in Ref.~\cite{Dimopoulos:2018xkm} including a gauge ensemble at the physical quark mass. \par

\section*{Acknowledgments}
We would like to thank all members of the European Twisted Mass Collaboration for the most enjoyable collaboration. Contributions by Krzysztof Cichy, Karl Jansen, Elena Garcia-Ramos, Chris Michael and Carsten Urbach at various stages of this project are gratefully acknowledged. The computer time for this project was made available to us by the John von Neumann-Institute for Computing (NIC) on the Jugene and Juqueen systems in J{\"u}lich. This project was funded by the DFG as a project in the SFB/TR 16 and the Sino-German CRC110. The gauge configurations used for this work have been generated by the European Twisted Mass Collaboration. The open source software packages tmLQCD~\cite{Jansen:2009xp} and Lemon~\cite{Deuzeman:2011wz} have been used.

\end{document}

%% file: newcommands.tex
\renewcommand{\l}{\left}
\renewcommand{\r}{\right}
\newcommand{\g}[1]{\gamma_{#1}} 
\newcommand{\tr}{\mathrm{tr}}
\newcommand{\diag}{\mathrm{diag}}  

\newcommand{\chiral}[1]{\mathring{#1}} 

\newcommand{\gev}{\,\mathrm{GeV}}

\newcommand{\mev}{\,\mathrm{MeV}}

\newcommand{\experiment}{\mathrm{exp}}
\newcommand{\phys}{\mathrm{phys}}
\newcommand{\stat}{\mathrm{stat}}
\newcommand{\sys}{\mathrm{sys}}
\newcommand{\tm}{\mathrm{tm}}

\newcommand{\Meta}{M_\eta}
\newcommand{\Metap}{M_{\eta^\prime}}

\newcommand{\chibar}{\overline{\chi}}